\documentclass[runningheads]{llncs}
\usepackage[T1]{fontenc}
\usepackage{graphicx}
\usepackage{cite}
\usepackage{amsmath,amssymb,amsfonts}
\usepackage{textcomp}
\def\BibTeX{{\rm B\kern-.05em{\sc i\kern-.025em b}\kern-.08em
    T\kern-.1667em\lower.7ex\hbox{E}\kern-.125emX}}

\usepackage[table]{xcolor}
\usepackage{orcidlink}
\usepackage{subcaption}
\usepackage{array}
\usepackage{tikz}
\usetikzlibrary{arrows.meta,positioning,fit,backgrounds}
\newcolumntype{P}[1]{>{\centering\arraybackslash}p{#1}}
\newcolumntype{M}[1]{>{\centering\arraybackslash}m{#1}}

\begin{document}
\title{PortLBM: A Portable Lattice Boltzmann Tool Leveraging SYCL on AMD, NVIDIA, and Intel GPUs}
\titlerunning{PortLBM}
\author{Alexander Strack \orcidID{0000-0002-9939-9044}
	\and Marcel Graf \orcidID{0009-0007-6308-1329}
	\and Alexander Van Craen \orcidID{0000-0002-3336-7226}
	\and Dirk Pflüger \orcidID{0000-0002-4360-0212}}
\authorrunning{A. Strack et al.}
\institute{Institute of Parallel and Distributed Systems, University of Stuttgart,\\ 70569 Stuttgart, Germany\\
	\email{\{alexander.strack, alexander.van-craen, dirk.pflueger\}@ipvs.uni-stuttgart.de}\newline
	\email{st172528@stud.uni-stuttgart.de}
}
\maketitle
\begin{abstract}
	The lattice Boltzmann method (LBM) is a well-established approach for simulating fluid flows at the mesoscopic scale.
	With the stagnation of Moore's law, high-performance computing has shifted toward GPU accelerators, necessitating programming models that ensure both portability and efficiency across diverse hardware platforms.

	We present PortLBM, an extensible portable LBM framework built on SYCL that integrates cross-platform GPU support with interactive real-time visualization.
	PortLBM supports diverse simulation scenarios ranging from Kármán vortex streets and wing flows to porous media, and is designed for easy extension with new algorithms and backends.
	As part of a performance portability study, we evaluate PortLBM on contemporary GPU architectures from NVIDIA, AMD, and Intel, examining the impact of three data layouts (stream, bundle, and collision) and four algorithmic variants on simulation throughput.

	Our results show that no single configuration achieves optimal performance across all GPU vendors, confirming the need for system-specific tuning.
	The stream layout maximizes bandwidth and performs best on the contemporary NVIDIA and Intel GPUs, while the bundle layout improves cache efficiency and excels on the AMD GPU.
	Two-lattice schemes achieve higher throughput while one-lattice schemes are preferable under memory constraints.
	Our work underscores the necessity for adaptable, portable LBM software in increasingly heterogeneous computing environments.

\end{abstract}

\section{Introduction}
Fluid flow simulations are central to many engineering and scientific applications, and the lattice Boltzmann method (LBM) is a popular mesoscopic alternative to traditional Navier–Stokes solvers.
As single-core scaling under Moore's law has slowed, GPU acceleration has become the dominant source of performance gains, producing a heterogeneous hardware landscape built by multiple vendors that requires programming models facilitating portable yet efficient implementations.
Various frameworks have been developed for this purpose, including OpenACC~\cite{Wienke2012_openacc}, OpenCL~\cite{Stone2010_opencl}, Kokkos~\cite{Edwards2014_kokkos}, and SYCL~\cite{SYCL2020}.
We choose SYCL for PortLBM due to its single-source ISO C\texttt{++} programming model, royalty-free open standard, and broad vendor support across Intel, AMD, and NVIDIA hardware through mature implementations such as DPC\texttt{++} and AdaptiveCpp.
Unlike Kokkos, SYCL aligns more closely with standard C\texttt{++} idioms and provides native portability for Intel GPU architectures.
The same PortLBM source compiles and runs unmodified across all three vendors, though achieving consistently high performance still requires the system-specific tuning we study in this work.

Regarding LBM, several factors influence performance across different architectures.
These include the collision operator, data layout, propagation scheme, and parallelization approach.
For the collision operator, the BGK operator~\cite{Bhatnagar1954_bkg} is widely used.
The second and third factors are investigated for CPUs in the work of Mattila et al.~\cite{Mattila2008_data_layouts}.
The authors compare four CPU-based LBM algorithms across three data layouts.
Motivated by this comparison, we explore the suitability of the three data layouts proposed in~\cite{Mattila2008_data_layouts} for modern GPU hardware.
Furthermore, we implement four portable LBM algorithms targeting GPUs, each focusing on a different optimization aspect, ranging from implementation complexity to memory consumption and performance, in our newly developed PortLBM tool, which also provides real-time visualization capabilities for non-stationary phenomena.
Specifically, the main contributions of this work include:
\begin{itemize}
	\item PortLBM: a portable, extensible LBM tool leveraging SYCL with integrated real-time visualization, supporting diverse simulation scenarios, and designed for straightforward extension with new algorithms and backends.
	\item The first application and evaluation of the Mattila et al.\ data layout taxonomy~\cite{Mattila2008_data_layouts} to modern GPU hardware, combining throughput benchmarks with hardware-counter profiling and theoretical kernel analysis.
    \item A cross-vendor performance portability study of four LBM algorithms across three data layouts (stream, bundle, collision) on contemporary NVIDIA, AMD, and Intel GPUs.
\end{itemize}

\begin{figure}[t]
	\centering
	\begin{subfigure}[t]{0.49\textwidth}
		\centering
		\includegraphics[width=\textwidth]{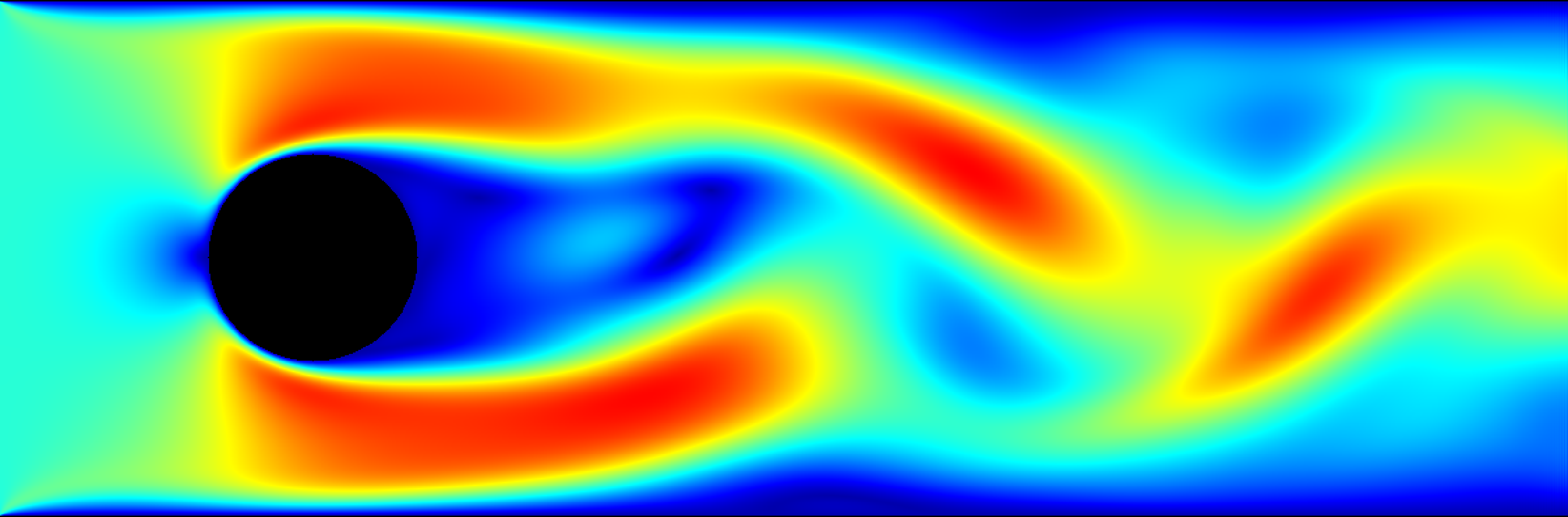}
		\caption{Vortex street around an obstacle}
		\label{fig:scenario-karman}
	\end{subfigure}
	\hfill
	\begin{subfigure}[t]{0.49\textwidth}
		\centering
		\includegraphics[width=\textwidth]{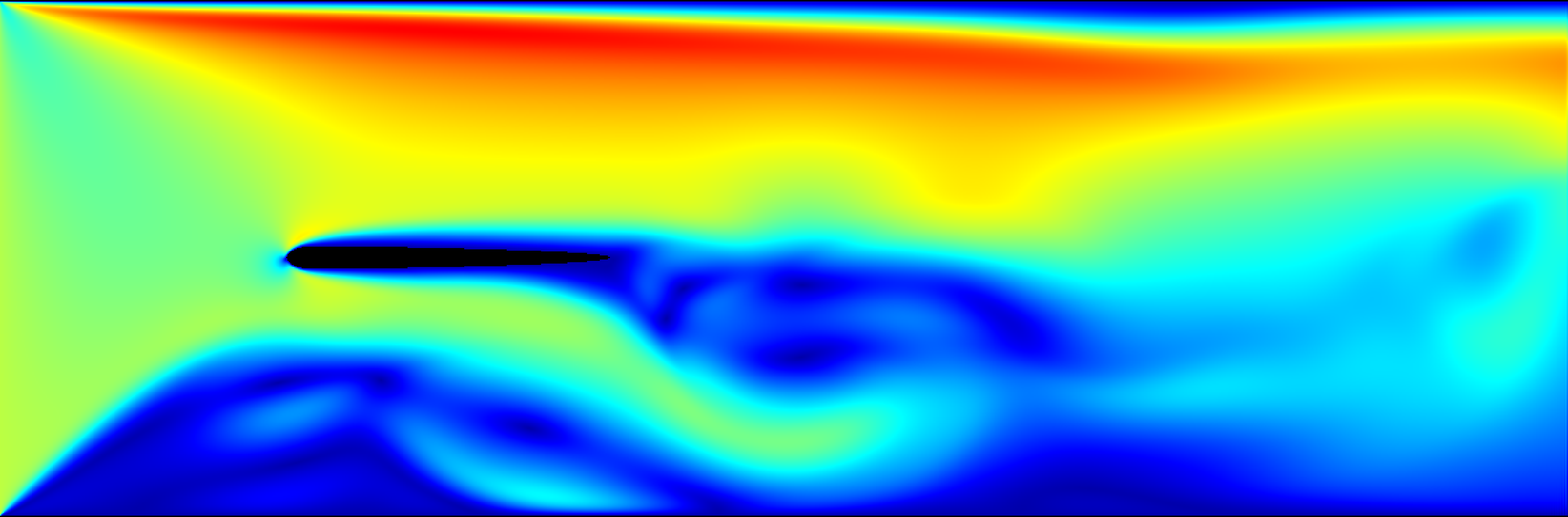}
		\caption{Wing flow at non-zero angle of attack}
		\label{fig:scenario-wing}
	\end{subfigure}
	\caption{Representative simulation scenarios supported by PortLBM.}
	\label{fig:intro-scenarios}
\end{figure}

PortLBM supports classical CFD benchmarks, including Hagen-Poiseuille flow and Kármán vortex streets around circular, square, and plate obstacles, as well as complex real-world cases such as wing flows at adjustable angles of attack, alternating obstacles, and porous media. Figure~\ref{fig:intro-scenarios} illustrates two representative examples.

The remainder of this work is structured as follows.
In Section \ref{sec:related_work}, we cover related work on portable LBM implementations.
Section \ref{sec:methods} introduces the theoretical background and the four algorithms.
In Section \ref{sec:software}, we describe the implementation of PortLBM and its software stack.
In Section \ref{sec:results}, we present and analyze the performance evaluation and profiling results.
Section \ref{sec:conclusion} concludes with a summary and future work directions.

\section{Related Work}\label{sec:related_work}

Large LBM software projects such as Palabos~\cite{Latt2021_palabos}, OpenLB~\cite{Krause2021_openlb}, and waLBerla~\cite{Godenschwager2013_walberla} have, at the time of writing, not yet fully ported their code base to portable programming models.
However, several smaller LBM projects have explored cross-platform programming models, which we review below.

In~\cite{Blair2015_openacc_lbm}, an existing D3Q15 solver using the Zou and He boundary conditions~\cite{Zou1995_bc_2d} is expanded with support for OpenACC.
Similarly, the authors of~\cite{Calore2016_openacc_lbm} use OpenACC to accelerate a special D2Q37 model based on a \emph{Structure of Arrays} (SoA) data layout and a \emph{two-step} algorithm.
In~\cite{Lee2019_openacc_lbm_harvey}, the Harvey codebase is ported to GPUs using CUDA and OpenACC.
Follow-up work added SYCL support to Harvey~\cite{Martin2023_sycl_lbm_harvey}.

The closest work to PortLBM is miniLB~\cite{Cosenza2025_minilb_sycl}, which also builds on SYCL and supports similar scenarios (driven cavity, Kármán vortex street, Hagen-Poiseuille flow, Taylor-Green vortices).
MiniLB uses the compressible D2Q9 model and implements a \emph{two-lattice} algorithm with a SoA data layout, writing output to VTK files.
In contrast, PortLBM uses the incompressible D2Q9I model~\cite{Zou1995_bc_2d}, supports three data layouts, reduces memory consumption with one-lattice variants, provides real-time visualization, and determines work-group size at runtime rather than compile time.

Another notable LBM code is the OpenCL-based FluidX3D library.
FluidX3D features low-precision lattice storage~\cite{Lehmann2022_fluidx3d_precision} and advanced streaming patterns~\cite{Lehmann2022_fluidx3d_scheme}.
The authors of~\cite{Verdier2020_kokkos_lbm} present LBM-scalay, which relies on Kokkos.
LBM-scalay supports a variety of lattice decompositions and uses a clustered SoA data layout.
In~\cite{Ding2023_sycl_kokkos}, the OpenLBMflow library is ported to SYCL and Kokkos.

None of the aforementioned works provides a comparative analysis of the impact of different data layouts proposed in~\cite{Mattila2008_data_layouts} on algorithmic performance, an evaluation whose relevance grows as architectures diverge across vendors.

\section{Methods}\label{sec:methods}

In this section, we cover the LBM foundations, the three data layouts described in~\cite{Mattila2008_data_layouts}, and the four algorithms implemented in PortLBM.

\subsection{Lattice Boltzmann Method}

The lattice Boltzmann equation (LBE), which underlies the LBM, is derived from the continuous Boltzmann equation via a Gauss-Hermite quadrature discretization of the velocity space~\cite{Schiller2008_lbm_theory}.
With the commonly used BGK relaxation~\cite{Bhatnagar1954_bkg} operator, the LBE can be formulated as
\begin{equation}
	f_i(\vec{r} + \vec{c_i}, t + 1) = f_i(\vec{r},t) - \frac{1}{\tau} \left[ f_i(\vec{r},t) - f_i^{eq}(\vec{r},t) \right],
	\label{eq:lattice-boltzmann-equation}
\end{equation}
for $i = 0, \dots, m-1$ with the position vector $\vec{r}$, the relaxation time $\tau$, and timestep $\Delta t = 1$ in lattice units, using the elementary velocities $\mathcal{C} = \{ \vec{c}_0, \dots, \vec{c}_{m-1}\} \subset \{-1,0,1\}^{n}$.

We use the D$2$Q$9$I model, which is specifically designed for incompressible fluids~\cite{Zou1995_bc_2d}.
The equilibrium distribution values $f_i^{eq}$ are given by
\begin{equation}
	f^{eq}_i = w_i \left[ \rho + 3 \Vec{v}  \cdot \vec{c_i}  + \frac{9}{2} (\vec{v}  \cdot \vec{c_i} )^2 - \frac{3}{2} \Vec{v}^2 \right]
	\label{eq:maxwell-boltzmann-d2q9i}
\end{equation}
with the velocity $\vec{v} = \sum_i f_i \cdot \vec{c_i}$, pressure $p = c_s^2 \cdot \rho \cdot \rho_0$, and weights $w_4 = \frac{4}{9}$, $w_1, w_3, w_5, w_7 = \frac{1}{9}$, $w_0, w_2, w_6, w_8 = \frac{1}{36}$.

For boundary treatment, PortLBM computes the equilibrium distribution via Dirichlet or Neumann conditions at the inlet, an adapted Zou and He condition~\cite{Zou1995_bc_2d} at the outlet, and a halfway bounce-back scheme for fluid-solid interactions.

\subsection{Data layouts}\label{sec:data_layout}

\begin{figure}[ht]
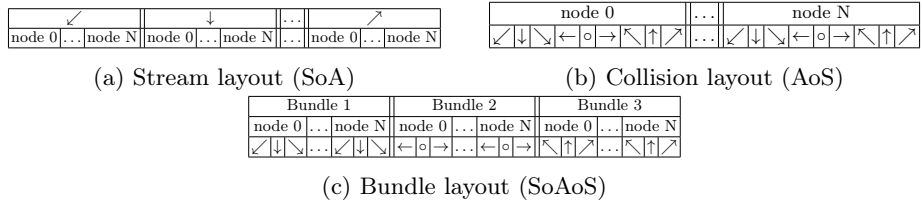

	\centering
	\begin{subfigure}[b]{0.48\textwidth}
		\resizebox{\textwidth}{!}{
			\begin{tabular}{|c|c|c||c|c|c||c||c|c|c|}
				\hline
				\multicolumn{3}{|c||}{$\swarrow$} &                                                                                        
				\multicolumn{3}{c||}{$\downarrow$}    %
				                                  & $\hdots$ &                                                                             %
				\multicolumn{3}{c|}{$\nearrow$}     %
				\\
				\hline
				node 0                            & $\hdots$ & node N & node 0 & $\hdots$ & node N & $\hdots$ & node 0 & $\hdots$ & node N \\
				\hline
			\end{tabular}
		}
		\caption{Stream layout (SoA)}
		\label{fig:stream}
	\end{subfigure}
	\hfill
	\begin{subfigure}[b]{0.48\textwidth}
		\resizebox{\textwidth}{!}{
			\begin{tabular}{|c|c|c|c|c|c|c|c|c||c||c|c|c|c|c|c|c|c|c|}
				\hline
				\multicolumn{9}{|c||}{node 0} &                                                                                                           
				$\hdots$                      &                                                                                                           %
				\multicolumn{9}{c|}{node N}     %
				\\
				\hline
				$\swarrow$                    & $\downarrow$ & $\searrow$ & $\leftarrow$ & $\circ$ & $\rightarrow$ & $\nwarrow$ & $\uparrow$ & $\nearrow$
				                              & $\hdots$     &
				$\swarrow$                    & $\downarrow$ & $\searrow$ & $\leftarrow$ & $\circ$ & $\rightarrow$ & $\nwarrow$ & $\uparrow$ & $\nearrow$ \\
				\hline
			\end{tabular}
		}
		\caption{Collision layout (AoS)}
		\label{fig:collision}
	\end{subfigure}
	\hfill
	\begin{subfigure}[b]{0.48\textwidth}
		\resizebox{\textwidth}{!}{
			\begin{tabular}{|c|c|c|c|c|c|c||c|c|c|c|c|c|c||c|c|c|c|c|c|c|}
				\hline
				\multicolumn{7}{|c||}{Bundle 1} &                                                                                         
				\multicolumn{7}{c||}{Bundle 2}  &                                                                                         %
				\multicolumn{7}{c|}{Bundle 3}     %
				\\
				\hline
				\multicolumn{3}{|c|}{node 0}
				                                & $\hdots$     &
				\multicolumn{3}{c||}{node N}    &
				\multicolumn{3}{c|}{node 0}
				                                & $\hdots$     &
				\multicolumn{3}{c||}{node N}    &
				\multicolumn{3}{c|}{node 0}
				                                & $\hdots$     &
				\multicolumn{3}{c|}{node N}
				\\
				\hline
				$\swarrow$                      & $\downarrow$ & $\searrow$    & $\hdots$ & $\swarrow$   & $\downarrow$ & $\searrow$    &
				$\leftarrow$                    & $\circ$      & $\rightarrow$ & $\hdots$ & $\leftarrow$ & $\circ$      & $\rightarrow$ &
				$\nwarrow$                      & $\uparrow$   & $\nearrow$    & $\hdots$ & $\nwarrow$   & $\uparrow$   & $\nearrow$      \\
				\hline
			\end{tabular}
		}
		\caption{Bundle layout (SoAoS)}
		\label{fig:bundle}
	\end{subfigure}
	\caption{Three supported data layouts in PortLBM proposed by~\cite{Mattila2008_data_layouts}.}
	\label{fig:data_layouts}
\end{figure}

The particle distribution values associated with each lattice node must be stored in a data structure.
Mattila et al.~\cite{Mattila2008_data_layouts} introduced three principal data layout strategies: stream, collision, and bundle.
The stream layout organizes the distribution values as an SoA (compare Figure \ref{fig:stream}).
This layout is particularly advantageous for the streaming step, as it enables threads processing consecutive nodes to access memory coalesced.
In contrast, the collision layout adopts an \emph{Array of Structures} (AoS) approach (see Figure \ref{fig:collision}).
The collision layout is well-suited for operations focused on individual nodes due to its superior cache efficiency.
The bundle layout represents a hybrid strategy by organizing data into an SoAoS (see Figure \ref{fig:bundle}).
It provides a middle ground between the bandwidth-focused stream layout and the cache-efficient collision layout.

\subsection{Algorithms}
Algorithms are derived by dividing Equation \ref{eq:lattice-boltzmann-equation} into a streaming and a collision step:

\begin{equation}
	\underbrace{f_i(\vec{r} + \vec{c_i}, t + 1)}_{\text{\textit{streaming} step}} = \underbrace{f_i(\vec{r},t) - \frac{1}{\tau} \left[ f_i(\vec{r},t) - f_i^{eq}(\vec{r},t) \right]}_{\text{\textit{collision} step}}.
	\label{eq:lbgk}
\end{equation}

There is a clear data dependency between the two steps, as each lattice node depends on the distribution values from its neighboring nodes' previous time steps.
In the literature, there are two main algorithmic approaches to handling this data dependency.
The \emph{two-step} approach iterates over the lattice twice.
In contrast, the \emph{two-lattice} approach uses a single combined embarrassingly parallel stream and collision step but relies on two lattices.

Classical streaming schemes are one-step pull and one-step push, where a node either pulls all required distribution values from neighboring nodes or pushes its own values to them.
Several more advanced streaming schemes have been proposed.
Current state-of-the-art approaches include the AA~\cite{Bailey2009_lbm_aa}, the Esoteric Twist~\cite{Geier2017_lbm_esoteric_twist}, and the Esoteric Pull~\cite{Lehmann2022_fluidx3d_scheme} schemes.

For PortLBM, we start with the basic \emph{two-lattice} approach, which relies on the one-step pull streaming scheme.
Each node pulls all incoming distribution values from its neighbors, combines the streaming and collision steps, and writes the result to a second, destination lattice.
Because the source lattice remains constant throughout a time step, all nodes are fully independent, eliminating the need for synchronization or buffering.

The simplest way of parallelizing the resulting algorithm using SYCL is the \texttt{sycl::range} kernel.
In this case, the work-group decomposition is entirely handled by the SYCL runtime, requiring no programmer knowledge of the underlying hardware.
The resulting algorithm is inherently portable and can be launched on any SYCL-compatible device without modification.
We refer to this algorithm as \emph{Linear Pull Two-Lattice} (LPTL).

Apart from linear \texttt{range}-based kernels, SYCL also supports explicit work-group decompositions with \texttt{nd\_range} kernels.
In this variant, the domain is treated as a two-dimensional grid and decomposed into equally shaped subdomains, each mapped to one work group.
Padding is applied so that the domain extents align with the work-group size, enabling coalesced memory accesses and giving the programmer direct control over work-group shape and size.
Since this algorithm still relies on the one-step pull scheme, we refer to it as \emph{Non-linear Pull Two-Lattice} (NPTL).

Both LPTL and NPTL require two full lattices in global memory.
To reduce memory consumption, our third variant avoids permanently allocating a second lattice.
Instead, each work item buffers all incoming distribution values for its node in SYCL private memory.
A local barrier then ensures that all work items within the work group have finished reading from the shared lattice before any writes occur, preventing data races within a work group.
Across work groups, a compact \emph{buffer grid} of additional nodes is woven into the domain to hold distribution values at the boundaries of adjacent subdomains.
This decouples subdomains without requiring global synchronization, keeping the total memory overhead well below that of a full second lattice.
We refer to this algorithm as \emph{Non-linear Pull One-Lattice} (NPOL).

\begin{sloppypar}
The fourth variant replaces the pull streaming scheme with the \emph{swap} scheme~\cite{Mattila2007_lbm_swap}.
Rather than copying incoming values from neighbors, each node actively exchanges its distribution value with the opposite value of the neighbor in each of the four positive directions.
In-flight values are buffered in SYCL local memory for exchanges within a work group.
The same buffer grid used in NPOL handles cross-subdomain interactions.
We refer to this algorithm as \emph{Non-linear Swap One-Lattice} (NSOL).
\end{sloppypar}

The four algorithms thus target distinct optimization aspects: simplicity (LPTL), peak throughput (NPTL), memory footprint (NPOL), and propagation scheme (NSOL).

\section{Implementation Details and Software Stack}\label{sec:software}

The PortLBM tool was designed to remain agnostic to the specific underlying LBM implementation.
Its GUI operates on instances of an abstract algorithm handler, which provides the central point of interaction between the simulation core and the visualization layer.
An abstract base class \texttt{AlgorithmHandler} defines the essential functionality for initializing and controlling simulations, as well as for accessing macroscopic observables.
SYCL integration is realized through the specialized class \texttt{SYCLAlgorithmHandler}, which internally manages a simulation algorithm bound to a particular simulation object.
Generally, PortLBM is designed to allow for easy extension, such as new algorithms or additional computation backends.
Importantly, the visualization framework itself interacts only with the generated simulation results, while remaining agnostic to internal data structures and implementation details.

PortLBM organizes the domain into equally sized work groups.
The user has two options at runtime.
Firstly, a reliable automatic decomposition is available, where the SYCL runtime chooses the work-group size: the domain is treated as a one-dimensional array iterated over in a linear \texttt{sycl::range<1>} kernel.
Secondly, a manual decomposition into a regular grid of subdomains of equal size and shape can be specified.
This is realized with the help of \texttt{sycl::nd\_range<2>} kernels.
PortLBM accepts a desired work-group size and selects a subdomain arrangement that minimizes the interface between them.
The choice of algorithm (LPTL, NPTL, NPOL, or NSOL) and data layout (stream, collision, or bundle) is likewise a runtime setting, specified via the \texttt{algorithm} and \texttt{dataLayout} fields of a JSON configuration file that is parsed at startup.
These settings, along with the work-group size, relaxation time, domain size, and scenario, can also be changed interactively from the GUI without recompilation.

\begin{sloppypar}
SYCL (pronounced \emph{sickle}) is a royalty-free, cross-platform programming model that lets developers write host and device code in standard ISO C\texttt{++} within a single-source model~\cite{SYCL2020}.
PortLBM is compiled with AdaptiveCpp (formerly hipSYCL)~\cite{Alpay2022_adaptive-cpp}, which interfaces with native hardware runtimes and supports NVIDIA, AMD, and Intel GPUs.
DPC\texttt{++}~\cite{Reinders2021_dpcpp} is Intel's open-source LLVM-based SYCL implementation and extends the standard with Intel-specific features.
\end{sloppypar}

Dear ImGui~\cite{dear-imgui} is a lightweight, self-contained C\texttt{++} GUI library supporting multiple rendering backends (DirectX12, Vulkan, OpenGL).
PortLBM uses it together with ImPlot~\cite{implot} for interactive, real-time plotting of macroscopic observables.
Both libraries were adapted to meet PortLBM's specific requirements.

\section{Results}\label{sec:results}

Hardware specifications are listed in Table~\ref{table:specs}.
We report lattice updates per second (LUPS) as the primary throughput metric.
Performance benchmarks (Section~\ref{sec:performance}) use the Hagen-Poiseuille scenario with 10 runs per configuration in double precision.
We report the median across these runs, with error bars in Figures~\ref{fig:best_per_layout} and~\ref{fig:best_per_algorithm} spanning the minimum to maximum observed value.
The domain sizes are $16000\times8000$ on the A30 and MI210, and $16000\times4000$ on the B580.
PortLBM was compiled with AdaptiveCpp, and the GUI was disabled during benchmarking.
Profiling runs (Section~\ref{sec:profiling}) use a thin vertical plate obstacle (Kármán vortex street). Obstacle and wing scenarios showed no notable performance difference compared to plain Hagen-Poiseuille, while a porous medium with randomly distributed solid nodes reduced throughput by $46\%$ on the A30 and $55\%$ on the MI210 due to reduced coalescing and increased bounce-back operations.
Extended work-group size comparisons are available in the accompanying thesis~\cite{Graf2025_lbm_sycl}.

\begin{table}[t]
	\centering
	\begin{minipage}[t]{.295\textwidth}
		\centering
		\Large
		\centering
		\resizebox{\textwidth}{!}{
			\begin{tabular}{|M{3cm}||M{3.5cm}|}
				\hline
				\cellcolor{gray!25}\textbf{CPU} & \cellcolor{gray!25}Dual AMD EPYC 9274F \\
				\hline\hline
				Cores                           & 48                                     \\
				\hline
				All-core boost                  & $3.1 \ \text{GHz}$                     \\
				\hline
				L1 cache                        & $64 \ \text{KB}$ / core                \\
				L2 cache                        & $1 \ \text{MB}$ / core                 \\
				\hline
				RAM                             & $384 \ \text{GB}$ DDR5                 \\
				\hline
				FP64                            & $3.11 \ \text{TFLOPS}$                 \\
				\hline
				Bandwidth                       & $460.8\ \text{GB}/\text{s}$            \\
				\hline
			\end{tabular}
		}
	\end{minipage}\hspace{.001\textwidth}
	\begin{minipage}[t]{.695\textwidth}
		\centering
		\huge
		\centering
		\resizebox{\textwidth}{!}{
			\begin{tabular}{|M{5cm}||M{5cm}|M{5cm}|M{5cm}|}
				\hline
				\cellcolor{gray!25}\textbf{GPU}        &
				\cellcolor{gray!25} NVIDIA A30         &
				\cellcolor{gray!25} AMD Instinct MI210 &
				\cellcolor{gray!25} Intel Arc B580
				\\
				\hline
				\hline
				Compute Units                          &
				$56$ SMs                               &
				$104$ CUs                              &
				20 Xe Cores 
				\\
				\hline
				L1 cache                               &
				$192 \ \text{KB}$ / SM                 &
				$16 \ \text{KB}$ / CU                  &
				$256 \ \text{KB}$ / Xe Core
				\\
				L2 cache                               &
				$24 \ \text{MB}$                       &
				$8 \ \text{MB}$                        &
				$18 \ \text{MB}$
				\\
				\hline
				VRAM                                   &
				$24 \ \text{GB}$ HBM2e                 &
				$64 \ \text{GB}$ HBM2e                 &
				$12 \ \text{GB}$ GDDR6
				\\
				\hline

				FP64                                   &
				$5.16 \ \text{TFLOPS}$                 &
				$22.63 \ \text{TFLOPS}$                &
				$1.71 \ \text{TFLOPS}$
				\\
				\hline
				Bandwidth                              &
				$933.1 \ \text{GB}/\text{s}$           &
				$1640 \ \text{GB}/\text{s}$            &
				$456.0 \ \text{GB}/\text{s} $
				\\
				\hline
			\end{tabular}
		}
	\end{minipage}
	\caption{Hardware specifications of CPU and GPUs.}
    \label{table:specs}
\end{table}

\subsection{Performance}\label{sec:performance}

\begin{figure}[t]
	\centering
	\begin{minipage}[t]{.47\textwidth}
		\centering
		\includegraphics[width=0.9\linewidth]{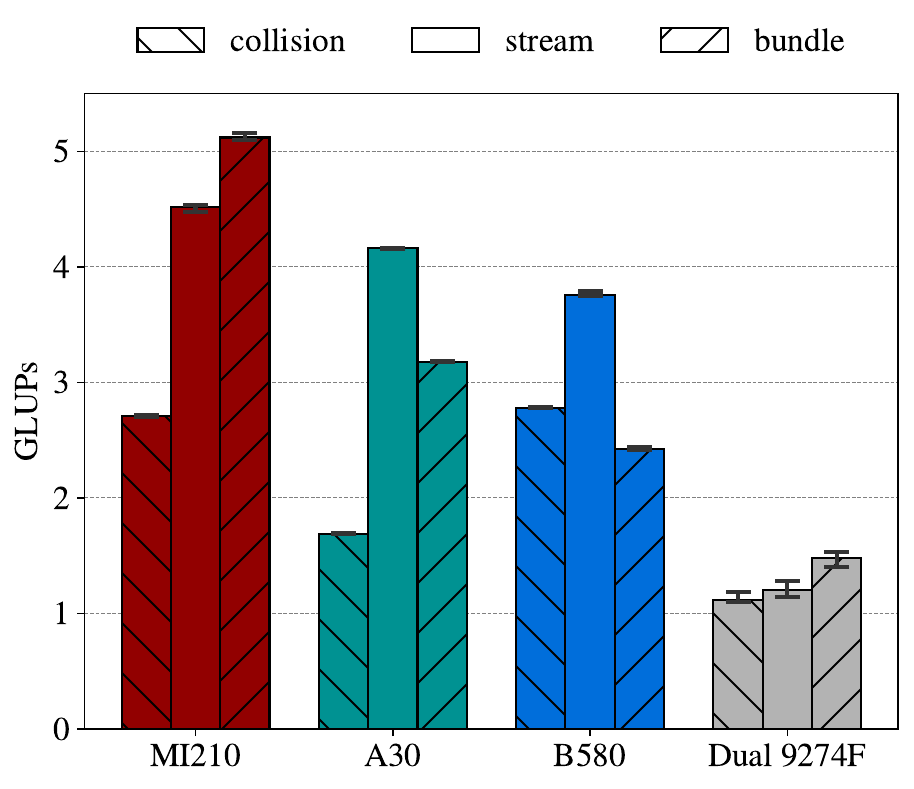}
		\caption{Lattice update performance of the best algorithm and work-group size for the three different data layouts on each system using double precision.}
		\label{fig:best_per_layout}
	\end{minipage}\hspace{.05\textwidth}
	\begin{minipage}[t]{.47\textwidth}
		\centering
		\includegraphics[width=0.9\linewidth]{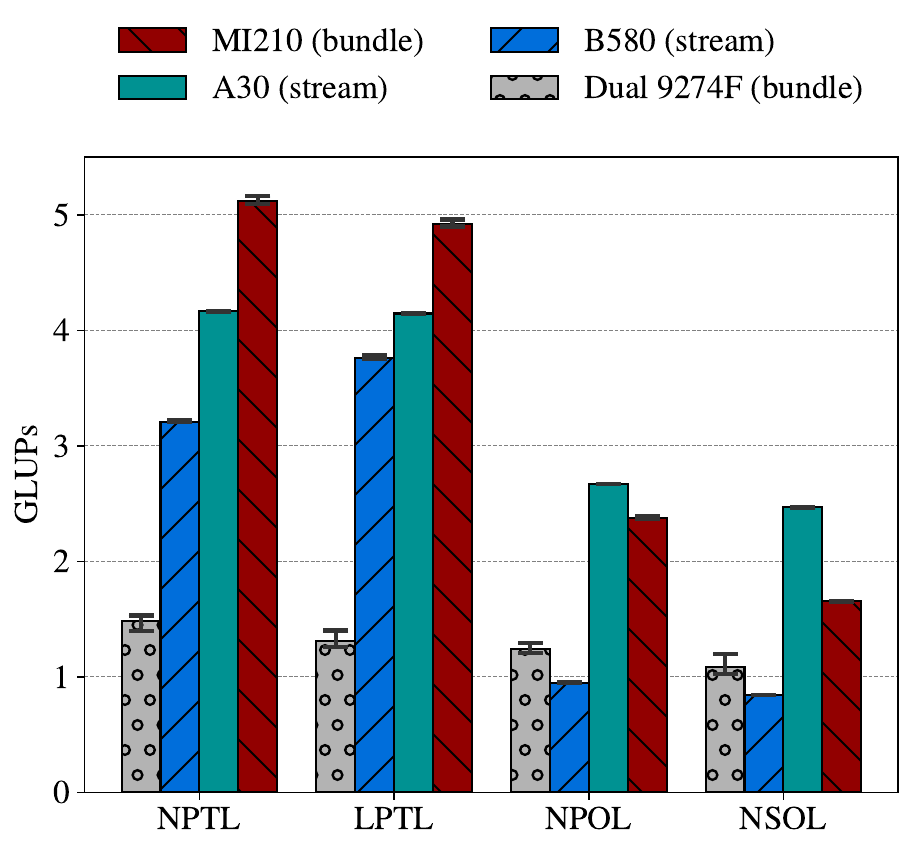}
		\caption{Lattice update performance comparison of the four different LBM algorithms using the best data layout and work-group size.}
		\label{fig:best_per_algorithm}
	\end{minipage}
\end{figure}

\begin{sloppypar}
In Figure \ref{fig:best_per_layout}, we present the LUPS performance using the best algorithm and work-group size for the three different data layouts.
On the reference AMD CPU, the bundle layout yielded the best performance.
The AMD CPU achieves $1.5$ GLUPS.
For the A30 and B580 GPUs with $4.16$ GLUPS and $3.76$ GLUPS, respectively, the stream layout yields the best performance.
On the AMD GPU, in contrast, the bundle layout yields the best performance, achieving $5.12$ GLUPS.
The collision layout is not competitive on the A30 ($1.73$ GLUPS) and MI210 ($2.71$ GLUPS).
On the B580, however, it achieves $2.82$ GLUPS, outperforming the bundle layout. 
This is owing to the B580's largest L1 cache per compute unit ($256$~KB/Xe Core vs.\ $192$~KB/SM on the A30 and $16$~KB/CU on the MI210), which enables better reuse of cached distribution values.
To further investigate the performance behavior, we evaluated the practical memory bandwidth of the three GPUs using an \href{https://github.com/ProjectPhysX/OpenCL-Benchmark}{OpenCL-Benchmark}\footnote{\url{https://github.com/ProjectPhysX/OpenCL-Benchmark} Last accessed: 2026-07-22}.
For coalesced read/write, the benchmark yields memory bandwidths of $807.33/914.99$~GB/s for the A30, $967.57/975.28$~GB/s for the MI210, and $584.93/474.28$~GB/s for the B580.
In contrast, for misaligned read/write, it yields memory bandwidths of $653.27/80.72$~GB/s for the A30, $1301.06/634.08$~GB/s for the MI210, and $893.07/398.42$~GB/s for the B580.
These results explain the comparable performance of the A30 and MI210 GPUs when using the stream layout optimized for coalesced memory accesses.
Furthermore, they help clarify the superior performance of the bundle layout on the AMD GPU.
Note that B580 results use a smaller domain ($16000\times4000$) and that its measured bandwidth exceeds the official specification.
Although this is consistent with reference values from the OpenCL-Benchmark authors, it should be treated as indicative rather than directly comparable to the A30 and MI210 results.
\end{sloppypar}

The actual performance gap between the four algorithms for their respective optimal data layouts and work-group sizes is shown in Figure \ref{fig:best_per_algorithm}.
For all systems except the B580 GPU, which performs best if the SYCL runtime decomposes the domain, the NPTL algorithm yields the best performance.
The performance improvements of the NPTL algorithm over the LPTL algorithm are $4.1\%$ on the MI210 GPU, $0.5\%$ on the A30 GPU, and $11.5\%$ on the CPU.
On the B580 GPU, we observe a $15.6\%$ decrease in performance.
The NPOL algorithm outperforms the NSOL algorithm by $8\%$ on other systems and up to $46\%$ on the MI210 GPU.
This performance delta is caused by the additional overhead introduced by a more complex streaming step and larger buffers, which make the NSOL algorithm not a viable option on our test hardware.

\subsection{Profiling}\label{sec:profiling}

Table~\ref{table:prof_algorithm} presents profiling results on the A30 GPU.
LPTL with the stream layout achieves the highest bandwidth ($82\%$ of peak) and best FLOPS, though only $14\%$ of peak compute.
NPTL improves cache hit rate and occupancy by operating on a regular work-group grid, at the cost of marginally lower bandwidth.
The two buffering algorithms (NPOL, NSOL) exhibit reduced occupancy and a zero L1 cache hit rate for the stream layout, because all distribution values must be loaded fresh on each update.
This result also explains the B580 performance drop for these algorithms (Figure~\ref{fig:best_per_algorithm}), as the large L1 cache goes unused.

Across the three NPTL data layouts, the stream variant achieves nearly twice the FLOPS and bandwidth of the bundle variant.
In contrast, the bundle variant yields significantly higher cache hit rates through its SoAoS structure.
The collision variant maximizes cache reuse, but its reduced effective bandwidth dominates, leading to the lowest throughput.
In Table~\ref{table:prof_algorithm}, the NSOL memory request count shows two values: total requests and, in parentheses, requests to SYCL local memory.

With the profiled values, we can analyze the efficiency of our kernels.
Take the combined stream and collision kernel of the NPTL algorithm.
It contains $9\cdot8~\text{B} + 1~\text{B}= 73~\text{B}$ worth of reading and $(9 + 4)\cdot8~\text{B}= 104~\text{B}$ worth of writing per lattice update, such that it has a code balance of $177~\text{B/LUP}$.
Given the specs of Table \ref{table:specs} we would expect a theoretical maximum of $\frac{933.1~\text{GB/s}}{177~\text{B/LUP}}\approx 5.27~\text{GLUPS}$ for the A30 GPU.
Taking the actual measured bandwidth, we should observe a performance of $\frac{730.38~\text{GB/s}}{177~\text{B/LUP}}\approx 4.13~\text{GLUPS}$.
The measured performance of $4.16$~GLUPS is within this range and may be slightly higher, as it also involves more lightweight boundary treatment kernels, indicating an efficient implementation in PortLBM.

Repeating this code balance roofline for the AMD GPU, we obtain a theoretical maximum of $\frac{1640~\text{GB/s}}{177~\text{B/LUP}}\approx 9.27~\text{GLUPS}$ for the MI210.
The best measured performance on the MI210, $5.12$~GLUPS with the bundle layout, corresponds to an effective bandwidth of $5.12~\text{GLUPS}\times177~\text{B/LUP}\approx 906.2~\text{GB/s}$.
This gap relative to the specified peak bandwidth quantifies the bandwidth the bundle layout trades away for cache locality.

Because our benchmarks indicate the B580's official bandwidth specification is under-reported, we omit a roofline analysis for that GPU.

\begin{table}[t]
	\centering
	\begin{minipage}[t]{\textwidth}
		\centering
		\large
		\resizebox{0.95\textwidth}{!}{
			\begin{tabular}{|M{2.5cm}||M{2cm}|M{2.5cm}|M{2.5cm}|M{2.5cm}|M{2cm}|M{2cm}|M{2.25cm}|}
				\hline
				Algorithm (Layout) & GFLOPS (FP64)              & Memory bandwidth (in $\text{GB/s}$) & Memory requests (in millions) & Arithmetic intensity       & L1 cache hit rate            & L2 cache hit rate            & Occupancy \textit{(limit)}                   \\
				\hline\hline
				Optimum            & $5161$                     & $933.1$                             & N/A                           & N/A                        & $100 \%$                     & $100 \%$                     & $100 \%$                                    \\
				\hline
				\hline
				NPTL (stream)      & $670.87$                   & $730.38$                            & $92.36$                       & \cellcolor{gray!20}$0.92$  & $14.41\%$                    & $71.63\%$                    & \cellcolor{gray!20}$63.13\%$ \textit{(75\%)} \\
				\hline
				NPTL (bundle)      & $345.37$                   & $377.82$                            & $92.36$                       & $0.91$                     & $64.04\%$                    & $78.92\%$                    & $60.34\%$ \textit{(75\%)}                    \\
				\hline
				NPTL (collision)   & $196.48$                   & $241.19$                            & $91.98$                       & $0.82$                     & \cellcolor{gray!20}$75.11\%$ & \cellcolor{gray!20}$86.73\%$ & $62.32\%$ \textit{(75\%)}                    \\
				\hline
				\hline
				LPTL (stream)      & \cellcolor{gray!20}$707.9$ & \cellcolor{gray!20}$769.93$         & \cellcolor{gray!20} $91.96$   & \cellcolor{gray!20} $0.92$ & $8.78\%$                     & $71.03\%$                    & $59.97\%$ \textit{(75\%)}                    \\
				\hline
				NPOL (stream)      & $513.14$                   & $572.20$                            & $99.98$                       & $0.90$                     & $0\%$                        & $69.42\%$                    & $45.90\%$ \textit{(50\%)}                    \\
				\hline
				NSOL (stream)      & $479.32$                   & $543.69$                            & $99.74$ $(26.40)$             & $0.88$                     & $0\%$                        & $68.65\%$                    & $43.20\%$ \textit{(50\%)}                    \\
				\hline
			\end{tabular}
		}
		\caption{Profiling results for the combined stream and collision kernel using the optimal work-group size per algorithm on an NVIDIA A30 GPU.}
		\label{table:prof_algorithm}
	\end{minipage}
\end{table}

\section{Conclusion and Outlook}\label{sec:conclusion}

In this work, we presented PortLBM, a novel portable LBM tool built on SYCL.
PortLBM supports a wide range of application scenarios and provides real-time visualization without additional software.
For PortLBM, we implemented three data layouts and four LBM algorithms, enabling system-specific performance tuning across different hardware platforms.
We performed an extensive performance evaluation and profiled multiple PortLBM configurations to assess how the insights of Mattila et al.~\cite{Mattila2008_data_layouts} translate to contemporary GPU architectures.

\begin{sloppypar}
Our results show that there is no one-size-fits-all configuration for performance-portable LBM code, as architectural designs (in particular, cache hierarchies and memory bandwidth) vary significantly across GPU vendors.
Specifically, the choice of data layout strongly influences performance.
The stream layout maximizes memory bandwidth but exhibits suboptimal cache utilization.
On both the NVIDIA and Intel GPUs, this layout provides the best overall performance.
In contrast, the bundle layout improves cache efficiency at the expense of bandwidth, making it the most effective layout on the AMD GPU tested here, a result driven by the specific cache design rather than a general AMD trait.
Consistent with \cite{Mattila2008_data_layouts}, the bundle layout delivers the highest performance on our reference CPU, although the performance differences are less significant than on the GPUs.
\end{sloppypar}

Regarding algorithmic variants, the \emph{two-lattice} algorithms implemented in PortLBM represent a trade-off: the linear variant is simple and efficient, whereas the non-linear variant achieves slightly higher performance and allows manual optimization of the domain decomposition via specifying the SYCL work-group size.
The \emph{one-lattice} variants, which rely on local buffering of distribution values, exhibit lower throughput but are advantageous when memory is constrained.
The swapping streaming scheme, taking advantage of work-group shared memory, does not yield any performance improvements.

Future work targets multi-GPU support with MPI integration and extension to three-dimensional problems.
Further directions include advanced streaming schemes (AA, Esoteric Twist), comparative studies across SYCL toolchains, and energy and performance modeling.

\begin{credits}
	\subsubsection{Reproducibility}

	The PortLBM tool is open source under the MIT license and is available on \href{https://github.com/SC-SGS/PortLBM}{GitHub}.\footnote{\url{https://github.com/SC-SGS/PortLBM} Last accessed: 2026-07-22}
	The software versions to reproduce the results presented in this work are: Clang 18.1.3, \text{AdaptiveCpp} v24.10, \text{Dear ImGui} v1.91.1, and \text{ImPlot} v0.17.

	\subsubsection{AI Usage Disclosure}

	Generative artificial intelligence (AI) tools, including Grammarly, ChatGPT, and Claude, were employed to enhance the clarity, grammar, and overall coherence of the manuscript. All technical content, data analyses, and research findings were conceived and developed independently by the authors. AI-assisted outputs were carefully reviewed, verified, and edited by the authors to ensure factual accuracy, interpretive rigor, and scholarly integrity. The final manuscript reflects the authors' original intellectual contributions and analytical work.
\end{credits}

\bibliographystyle{splncs04}
\bibliography{main}

@article{Calore2016_openacc_lbm,
  author       = {Calore, Enrico and Gabbana, Alessandro and Kraus, Jiri and others},
  title        = {Performance and Portability of Accelerated Lattice Boltzmann Applications with OpenACC},
  journal      = {Concurrency Comput. Pract. Exper.},
  volume       = {28},
  number       = {12},
  pages        = {3485--3502},
  year         = {2016},
}

@article{Cosenza2025_minilb_sycl,
title = {miniLB: Benchmarking Lattice Boltzmann simulations on AMD, Intel, and NVIDIA GPUs},
journal = {Future Gener. Comput. Syst.},
volume = {175},
pages = {108032},
year = {2026},
issn = {0167-739X},
author = {Biagio Cosenza and Luigi Crisci and Giorgio Amati and others},
}

@inproceedings{Ding2023_sycl_kokkos,
  author={Ding, Yue and Xu, Chuanfu and Qiu, Haozhong and others},
  booktitle={2023 ISPA/BDCloud/SocialCom/SustainCom}, 
  title={Evaluating Performance Portability of SYCL and Kokkos: A Case Study on LBM Simulations}, 
  year={2023},
  volume={},
  number={},
  pages={328-335},
}

@inproceedings{Blair2015_openacc_lbm,
author = {Blair, Stu and Albing, Carl and Grund, Alexander and others},
title = {Accelerating an MPI Lattice Boltzmann code using OpenACC},
year = {2015},
publisher = {ACM},
address = {New York, NY, USA},
booktitle = {WACCPD '15},
articleno = {3},
numpages = {9},
location = {Austin, Texas}
}

@article{Lee2019_openacc_lbm_harvey,
title = {Performance portability study for massively parallel computational fluid dynamics application on scalable heterogeneous architectures},
journal = {J. Parallel Distrib. Comput.},
volume = {129},
pages = {1-13},
year = {2019},
issn = {0743-7315},
author = {Seyong Lee and John Gounley and Amanda Randles and others},
}

@article{Lehmann2022_fluidx3d_precision,
  title = {Accuracy and performance of the lattice Boltzmann method with 64-bit, 32-bit, and customized 16-bit number formats},
  author = {Lehmann, Moritz and Krause, Mathias J. and Amati, Giorgio and others},
  journal = {Phys. Rev. E},
  volume = {106},
  issue = {1},
  pages = {015308},
  numpages = {28},
  year = {2022},
  month = {Jul},
  publisher = {American Physical Society},
}

@article{Lehmann2022_fluidx3d_scheme,
author = {Lehmann, Moritz},
title = {Esoteric Pull and Esoteric Push: Two Simple In-Place Streaming Schemes for the Lattice Boltzmann Method on GPUs},
journal = {Computation},
volume = {10},
year = {2022},
number = {6},
articleno = {92},
issn = {2079-3197},
}

@inproceedings{Martin2023_sycl_lbm_harvey,
title = {Performance Evaluation of Heterogeneous GPU Programming Frameworks for Hemodynamic Simulations},
author = {Martin, Aristotle and Liu, Geng and Ladd, William and others},
year = {2023},
publisher = {ACM},
address = {New York, NY, USA},
booktitle = {SC-W '23},
pages = {1126–1137},
numpages = {12},
location = {Denver, CO, USA}
}

@article{Verdier2020_kokkos_lbm,
title = {Performance portability of lattice Boltzmann methods for two-phase flows with phase change},
journal = {Comput. Methods Appl. Mech. Eng.},
volume = {370},
pages = {113266},
year = {2020},
issn = {0045-7825},
author = {Werner Verdier and Pierre Kestener and Alain Cartalade}
}

@phdthesis{
    Schiller2008_lbm_theory,
    author = {Ulf Daniel Schiller},
    title = {Thermal fluctuations and boundary conditions in the lattice Boltzmann method},
    school = {Johannes Gutenberg-Universität Mainz},
    year = {2008},
    address = {Mainz},
}

@article{Zou1995_bc_2d,
author={Zou, Qisu
and Hou, Shuling
and Chen, Shiyi
and others},
title={A improved incompressible lattice Boltzmann model for time-independent flows},
journal={J. Stat. Phys.},
year={1995},
month={Oct},
day={01},
volume={81},
number={1},
pages={35-48},
issn={1572-9613},
}

@article{Mattila2008_data_layouts,
title = {Comparison of implementations of the lattice-Boltzmann method},
author={Mattila, Keijo and Hyv{\"a}luoma, Jari and Timonen, Jussi and others},
journal = {Comput. Math. Appl.},
volume = {55},
number = {7},
pages = {1514-1524},
year = {2008},
issn = {0898-1221},
}

@article{Mattila2007_lbm_swap,
title = {An efficient swap algorithm for the lattice Boltzmann method},
journal = {Comput. Phys. Commun.},
volume = {176},
number = {3},
pages = {200-210},
year = {2007},
issn = {0010-4655},
author = {Keijo Mattila and Jari Hyväluoma and Tuomo Rossi and others},
}

@article{Bhatnagar1954_bkg,
  title = {A Model for Collision Processes in Gases. I. Small Amplitude Processes in Charged and Neutral One-Component Systems},
  author = {Bhatnagar, P. L. and Gross, E. P. and Krook, M.},
  journal = {Phys. Rev.},
  volume = {94},
  issue = {3},
  pages = {511--525},
  numpages = {0},
  year = {1954},
  month = {May},
  publisher = {American Physical Society},
}

@inproceedings{Alpay2022_adaptive-cpp,
author = {Alpay, Aksel and Soproni, B\'{a}lint and W\"{u}nsche, Holger and others},
title = {Exploring the possibility of a hipSYCL-based implementation of oneAPI},
year = {2022},
publisher = {ACM},
address = {New York, NY, USA},
booktitle = {IWOCL '22},
articleno = {10},
numpages = {12},
location = {Bristol, United Kingdom}
}

@inproceedings{Wienke2012_openacc,
author="Wienke, Sandra
and Springer, Paul
and Terboven, Christian
and others",
title="OpenACC --- First Experiences with Real-World Applications",
booktitle="Euro-Par 2012 Parallel Processing",
year="2012",
publisher="Springer",
address="Berlin, Heidelberg",
pages="859--870",
isbn="978-3-642-32820-6"
}

@article{Edwards2014_kokkos 
,
title = {Kokkos: Enabling manycore performance portability through polymorphic memory access patterns},
journal = {J. Parallel Distrib. Comput.},
volume = {74},
number = {12},
pages = {3202-3216},
year = {2014},
issn = {0743-7315},
author = {H. {Carter Edwards} and Christian R. Trott and Daniel Sunderland},
}

@article{Stone2010_opencl,
  author={Stone, John E. and Gohara, David and Shi, Guochun},
  journal={Comput. Sci. Eng.}, 
  title={OpenCL: A Parallel Programming Standard for Heterogeneous Computing Systems}, 
  year={2010},
  volume={12},
  number={3},
  pages={66-73},
}

@misc{SYCL2020,
  author       = {Khronos Group},
  title        = {SYCL 2020 Specification},
  year         = {2021},
  howpublished = {\url{https://registry.khronos.org/SYCL/specs/sycl-2020/html/sycl-2020.html}},
  note         = {Last accessed: 2026-07-22}
}

@article{Latt2021_palabos,
author = {Jonas Latt and Orestis Malaspinas and Dimitrios Kontaxakis and others},
title = {Palabos: Parallel Lattice Boltzmann Solver},
journal = {Comput. Math. Appl.},
volume = {81},
pages = {334-350},
year = {2021},
issn = {0898-1221},
}

@article{Krause2021_openlb,
title = {OpenLB—Open source lattice Boltzmann code},
journal = {Comput. Math. Appl.},
volume = {81},
pages = {258-288},
year = {2021},
issn = {0898-1221},
author = {Mathias J. Krause and Adrian Kummerländer and Samuel J. Avis and others},
}

@inproceedings{Godenschwager2013_walberla,
author = {Godenschwager, Christian and Schornbaum, Florian and Bauer, Martin and others},
title = {A framework for hybrid parallel flow simulations with a trillion cells in complex geometries},
year = {2013},
publisher = {ACM},
address = {New York, NY, USA},
booktitle = {SC '13},
articleno = {35},
numpages = {12},
location = {Denver, Colorado}
}

@misc{dear-imgui,
  author       = {Omar Cornut},
  title        = {{Dear ImGui}},
  howpublished = {\url{https://github.com/ocornut/imgui}},
  note         = {Version 1.91.1, Last accessed: 2026-07-22},
  year         = {2014}
}

@misc{implot,
  author       = {Evan Pezent},
  title        = {{ImPlot}},
  howpublished = {\url{https://github.com/epezent/implot/}},
  note         = {Last accessed: 2026-07-22},
  year         = {2020}
}

@article{Geier2017_lbm_esoteric_twist,
author = {Geier, Martin and Schönherr, Martin},
title = {Esoteric Twist: An Efficient in-Place Streaming Algorithmus for the Lattice Boltzmann Method on Massively Parallel Hardware},
journal = {Computation},
volume = {5},
year = {2017},
number = {2},
articleno = {19},
issn = {2079-3197},
}

@inproceedings{Bailey2009_lbm_aa,
  author={Bailey, Peter and Myre, Joe and Walsh, Stuart D.C. and others},
  booktitle={2009 ICPP}, 
  title={Accelerating Lattice Boltzmann Fluid Flow Simulations Using Graphics Processors}, 
  year={2009},
  volume={},
  number={},
  pages={550-557},
}

@book{Reinders2021_dpcpp,
  title={Data Parallel C++: Mastering DPC++ for Programming of Heterogeneous Systems using C++ and SYCL},
  author={Reinders, James and Ashbaugh, Ben and Brodman, James and others},
  year={2021},
  publisher={Apress},
}

@misc{Graf2025_lbm_sycl,
    author = {Marcel Graf},
    title = {{Real-time visualized and GPU-accelerated lattice Boltzmann simulations}},
    pages = {125},
    type = {Bachelor Thesis},
    number = {66},
    month = {April},
    year = {2025},
    language = {English},
    ee = {https://doi.org/10.18419/opus-17193},
    howpublished = {University of Stuttgart}
 }

\end{document}